\documentclass[aps,prl,twocolumn,amsmath,amssymb,groupedaddress,showpacs,showkeys]{revtex4-1}
\usepackage{graphicx}
\begin{document}


\title{Unravelling the puzzling intermediate states in the Biham-Middleton-Levine traffic model}

\author{L.E. Olmos}
\email[]{leolmoss@unal.edu.co}
\affiliation{Simulation of Physical Systems Group, CeiBA-Complejidad, Physics Department, National University of Colombia, Cra. 30 \# 45-03, Ed.404, Of, 348, Bogot\'a D.C., Colombia}
\author{J. D. Mu\~noz}
\email[]{jdmunozc@unal.edu.co}
\affiliation{Simulation of Physical Systems Group, CeiBA-Complejidad, Physics Department, National University of Colombia, Cra. 30 \# 45-03, Ed.404, Of, 348, Bogot\'a D.C., Colombia}


\date{\today}

\begin{abstract}
The Biham-Middleton-Levine (BML) traffic model, a cellular automaton with east-bound and north-bound cars moving by turns on a square lattice, has been an underpinning model in the study of collective behaviour by cars, pedestrians and even internet packages. Contrary to initial beliefs that the model exhibits a sharp phase transition from freely flowing to fully jammed, it has been reported that it shows intermediate stable phases, where jams and freely flowing traffic coexist, but  there is no clear understanding of their origin. Here, we analyze the model as an anisotropic system with a preferred fluid direction (north-east) and find that it exhibits two differentiated phase transitions: either if  the system is longer in the flow direction (longitudinal) or perpendicular to it  (transversal).  The critical densities where these transitions occur enclose the density interval of intermediate states and can be approximated by mean-field analysis, all derived from the anisotropic exponent relating the longitudinal and transversal correlation lengths. Thus, we arrive to the interesting result that the puzzling intermediate states in the original model are just a superposition of these two different behaviours of the phase transition, solving by the way most mysteries behind the BML model, which turns to be a paradigmatic example of such anisotropic critical systems.
\end{abstract}


\maketitle


In the recent urbanization era, society faces an inevitably increase of traffic congestion, turning its attention onto urban road networks. Nowadays is widely assumed that a proper understanding of the mechanisms leading jamming processes is indispensable for improving the efficiency of transportation systems. The Biham-Middleton-Levine (BML) model \cite{biham92}  is, perhaps, the simplest traffic cellular automaton able to exhibit self-organization, pattern formation and phase transitions \cite{biham92, tadaki, kertesz, gupta}. Hence, much extensive research on flux an collective behavior has been based on it, not just for car traffic (see \cite{lightstrategy, kinetic} and Refs. in \cite{ nagatani,chowd}) but also for pedestrian \cite{Hao,hilhorst} and information packages on the Internet \cite{internet}.
For more than a decade, it has been believed that at a certain critical car density $\rho_c$, the system exhibits what seems to be a first order phase transition between two phases: a free-flowing phase, where all cars move freely at all time steps (the average velocity of cars $v$$=$$1$) and a completely jamming phase, where no car moves at all ($v$$=$$0$). The value of $\rho_c$ decreases with increasing system size, possibly reaching the value $\rho_c$$=$$0$ as the system size approaches infinity. Then, it was thought that the BML model would be similar to other well known systems in statistical physics exhibiting phase transitions, e.g. percolation.
\begin{figure}[b]
\includegraphics[width=0.3\textwidth]{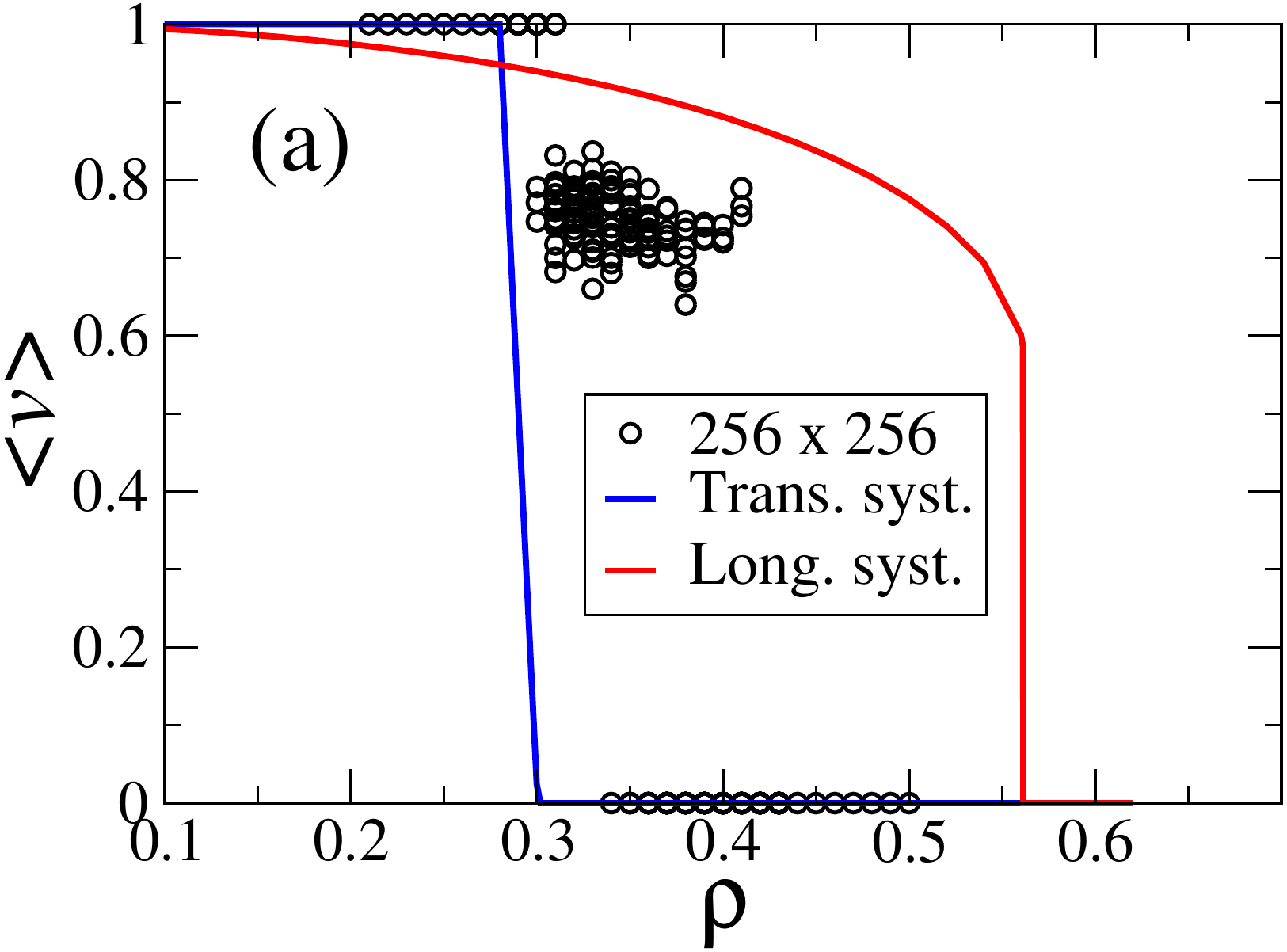}
\includegraphics[width=0.127\textwidth]{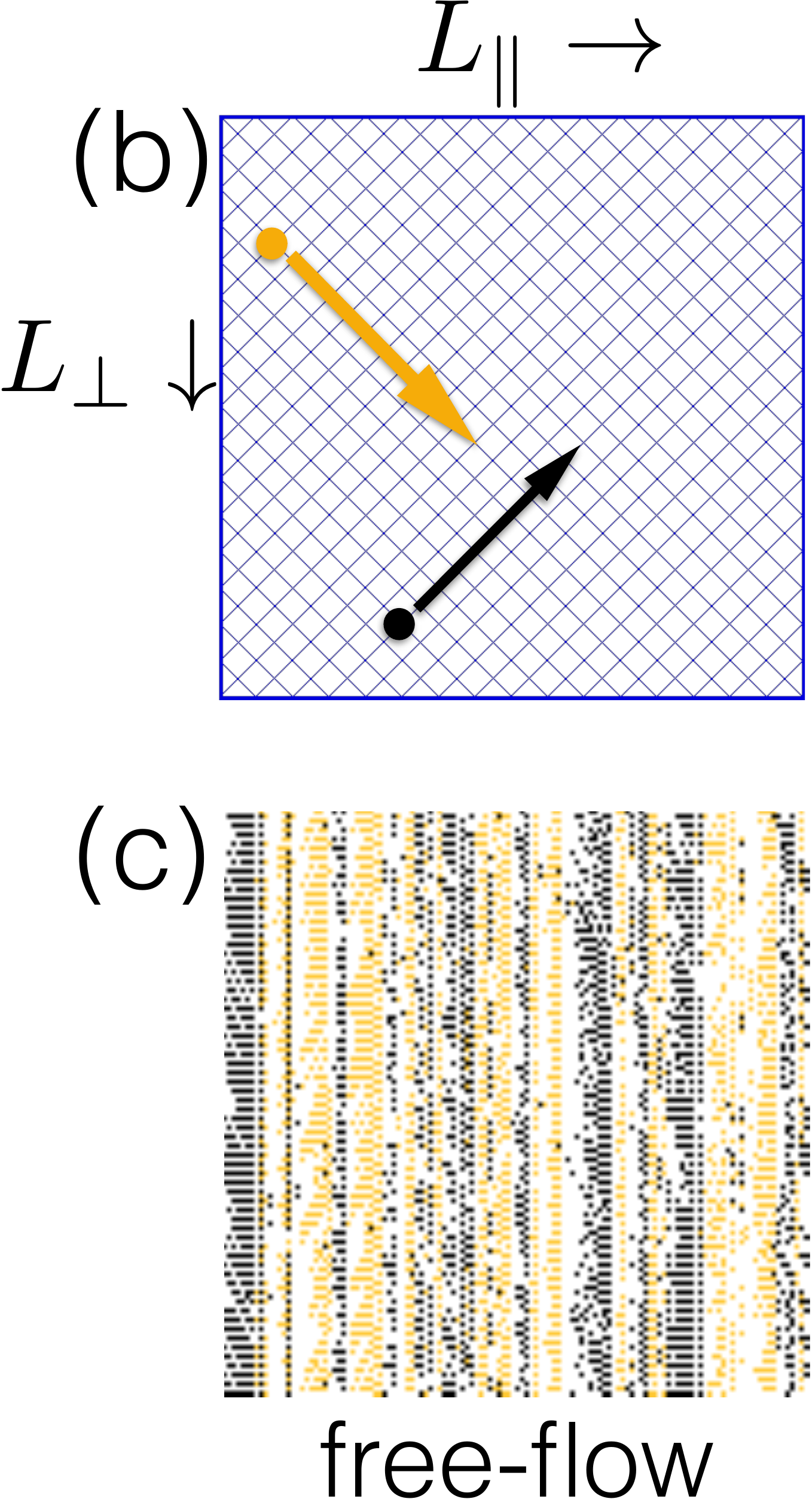}
\hspace{2cm}\includegraphics[width=0.435\textwidth]{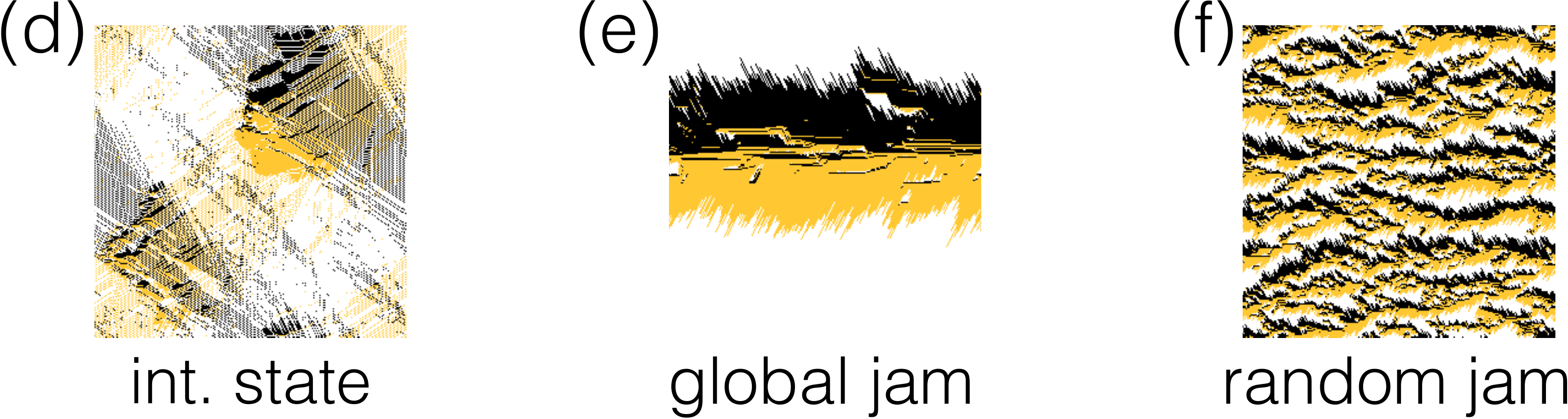}
\caption{(color online) (a) Average velocity $\left< v\right>$ vs density $\rho$ (circles) for a BML model on a $45^o$-rotated $256\times256$ lattice (as depicted in (a),  with east-bound cars (brown) and north-bound cars (black)). The mean-field approach for longitudinal systems (red line) and the numerical prediction for an infinite transverse system (blue line) are also included for comparison. Typical configurations for free flow (c), intermediate states (d), one global jam (e) and random jams (f) are also included.}
\label{FigIntermediateStates}
\end{figure}
However, all these conventional beliefs started being reconsidered, since W.K. Young\cite{wang} and R. M. D'Souza \cite{raissa1,raissa2} realized the existence of intermediate stable phases where free-flowing and jamming phases coexist(see Fig.\ref{FigIntermediateStates}). The structure of these states is highly regular, with jams' wavefronts  moving through  freely flowing traffic in a wide density region, and the value of the average velocity ($0$$<$$v$$<1$)  is extremely sensitive to the \text{aspect ratio} of the underlying lattice. Thus, instead of a phase transition as a function of car density, the system would exhibit two bifurcation points, limiting a region where intermediate phases would coexist between the two conventional phases above. Even though these states have been described and their asymptotic speeds have been predicted \cite{raissa1}, the exact locations of the bifurcation points are very difficult to determine and no one knows what truly happens as the system size goes to infinity. Moreover, the origin of the intermediate states remain as a unsolved puzzle.

In this Letter we focus on the clear existence of a preferred direction in the model dynamics. Most previous studies have overlooked this feature, despite \-- as we will show \-- it is the key  for unraveling  the puzzling intermediate states of the BML model. The presence of anisotropy has been fundamental in the analysis of force networks in granular matter\cite{ostojic, jamming} and flocking in collective animal behavior \cite{vicsek}, and classical theoretical models with this features are the next-nearest-neighbor Ising model (ANNNI)\cite{henkel} and the driven lattice gas model\cite{lattices}, and it should be naturally reflected in the phase transition of BML, making it in principle more akin to the anisotropic equivalent in percolation: directed percolation (DP) \cite{hinrich, nonequi}. By performing a  direct anisotropic scaling analysis on the BML model phase transition\cite{binder,redner}, we  found that the jamming process behaves distinctively as two separated phase transitions along different directions, namely, whether the system is longer in the direction of traffic flow (longitudinal system) or in the transversal direction (transversal system). Our main result is that the puzzling intermediate states in the BML model on square lattices emerge just as a superposition of these two different transitions, solving \-- by the way \-- most mysteries behind the BML model. 

{\it Model -} 
The original BML model considers two types of cars: east-bound (brown) and north-bound (black), moving on a two-dimensional square lattice with periodic boundary conditions. Each lattice site is in one of three states: empty, occupied by a brown car, or occupied by a black one. The cars are initially randomly distributed over the lattice sites with spatial density $\rho$ (usually taken to be the same, $\rho/2$, for both north- and eastbound cars). The fully deterministic dynamics is as follows: On even (odd) steps, all eastbound (northbound) cars synchronously attempt to advance one lattice site toward the east (north). If the site eastward (northward) of a car is currently empty, it advances. Otherwise, it remains stationary. The system exhibits, therefore, a preferred north-east direction for the flux. 

To study the system as an anisotropic one, we explicitly rotated 45 degrees the lattice, so we could control the system lengths along the longitudinal ($L_\parallel$) and transversal ($L_\perp$) directions to the car flow. Although this rotation changes the boundary conditions,  the system still exhibits the same three phases observed in the original BML model (see Fig.\ref{FigIntermediateStates}). In anisotropic systems, clusters show different correlation lengths along the longitudinal and transversal directions, $\xi_\parallel$ and $\xi_\perp$, respectively, which scale with different exponents as $ \xi_\parallel \sim(\rho-\rho_c)^{-\frac{1}{\nu_\parallel}}\quad \text{and}\quad \xi_\perp \sim (\rho-\rho_c)^{-\frac{1}{\nu_\perp}}$ \cite{binder,hinrich}.
An anisotropy exponent, relating the different scaling of the two correlation lengths, is defined as the ratio $\theta$$=$$\frac{\nu_\parallel}{\nu_\perp}$. According to
  \cite{binder,redner,williams}, when the longitudinal and transversal lengths are related by the constraint $L_\perp\sim L_{\parallel}^{\frac{1}{\theta}}$, the system behaves as it were effectively isotropic, and standard finite-size scaling (FSS) theory applies again for all percolation quantities, just in terms of the length scale $L_\parallel$.  Especially the transition width and the percolation threshold scale like 
\begin{equation}
\Delta(L_\parallel, L_\parallel^{\frac{1}{\theta}})\sim L_\parallel^{-\frac{1}{\nu_\parallel}} \text{   and   } \left|\rho_c - \left< \rho_c(L_\parallel,L_\parallel^{\frac{1}{\theta}})\right>\right| \sim L_\parallel^{-\frac{1}{\nu_\parallel}} .
\label{scale}
\end{equation}
The exponent $\theta$ can be estimated numerically from the fact that, close to the critical point, the two correlations lengths must be related by $\xi_\parallel\sim\xi_\perp^{\theta}$. The symbol $\langle\rangle$ denotes averages over final jammed configurations starting from different random initial conditions. With this theory in mind, let us define the parallel (perpendicular) spatial correlation function \cite{tadaki} as
\begin{equation}
G_{\parallel (\perp)}(\vec{r'})=\frac{1}{N}\left\langle\sum_{\vec{r}} \sigma(\vec{r})\cdot\sigma(\vec{r}+\vec{r'})\right\rangle \quad,
\end{equation}
where $\sigma(\vec{x})$$=$$1(0)$ if the site with position $\vec{x}$ is occupied(empty), N is the total number of cars and $\vec{r'}$ is a vector in the direction $\parallel$ ($\perp$) you want to compute the correlation function along. The correlation functions are fitted with exponentials $G_{\parallel(\perp)}\propto\exp(-r/\xi_{\parallel(\perp)})$ to estimate $\xi_{\parallel(\perp)}$. 
\begin{figure}[t]
\includegraphics[width=0.36\textwidth]{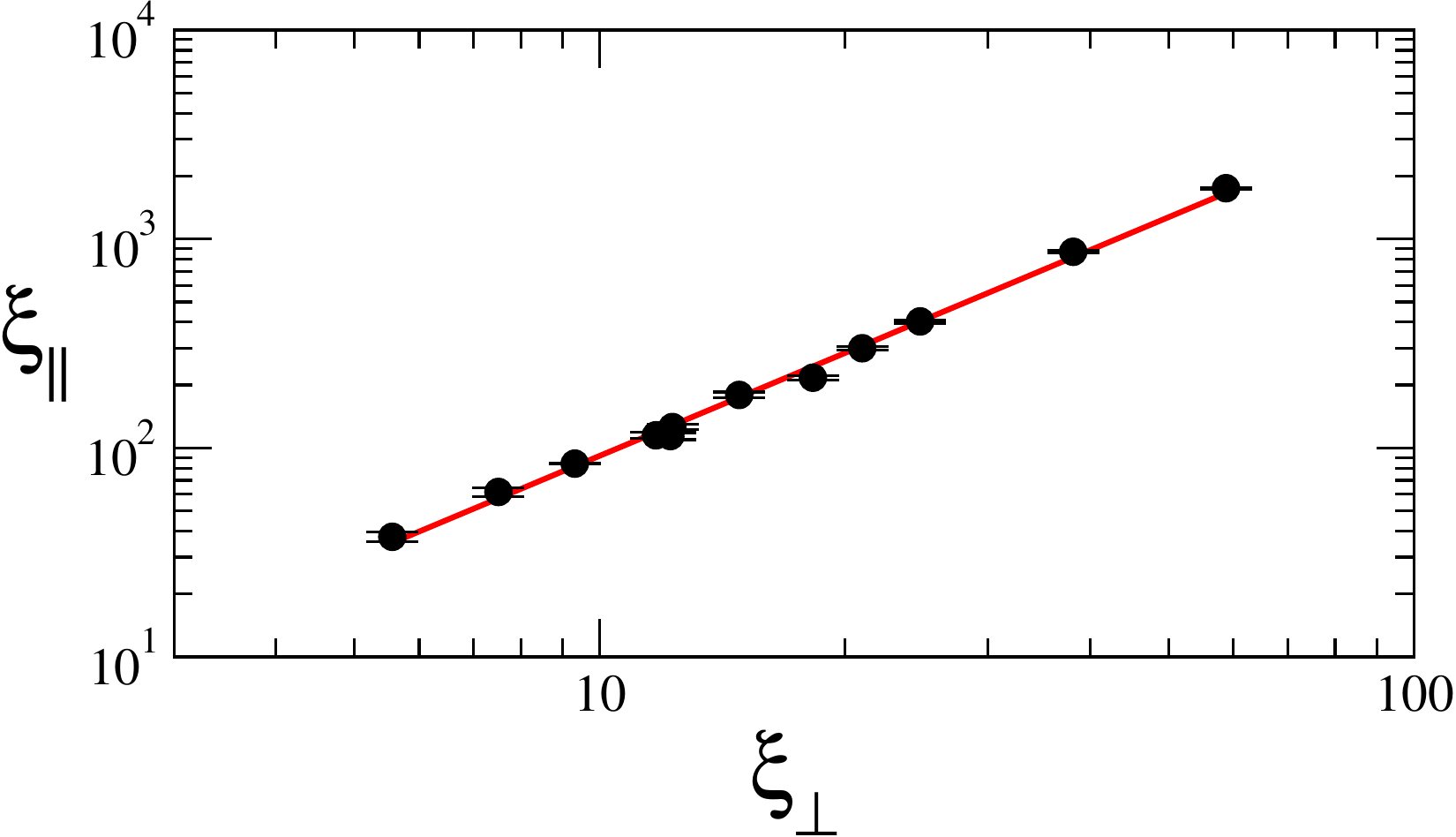}
\caption{(color online). Longitudinal $\xi_\parallel$ and transversal $\xi_\perp$ correlation lengths from final configurations at densities $\rho$ in the range [0.49-0.54] for square lattices of different sizes. The power-law fit gives $\theta$$=$$1.64(3)$ for the anisotropy exponent.}
\label{Corr}
\end{figure}
Figure \ref{Corr} presents the correlations lengths computed from final configurations of the original BML model for different square lattice sizes at densities close to the threshold transition, averaging over 50 configurations for each point. A linear regression yields an estimate of the anisotropy exponent $\theta$$=$$1.64\pm0.03$ (Here and everywhere the error bars are 1$\sigma$). 

With the exponent $\theta$ in hand, we studied the phase transition for both longitudinal and transversal systems. Longitudinal ones ran on lattices with sizes $ L_{\parallel}^{\frac{1}{\theta}}\times L_{\parallel}$.
\begin{figure*}[]
\includegraphics[width=0.36\textwidth]{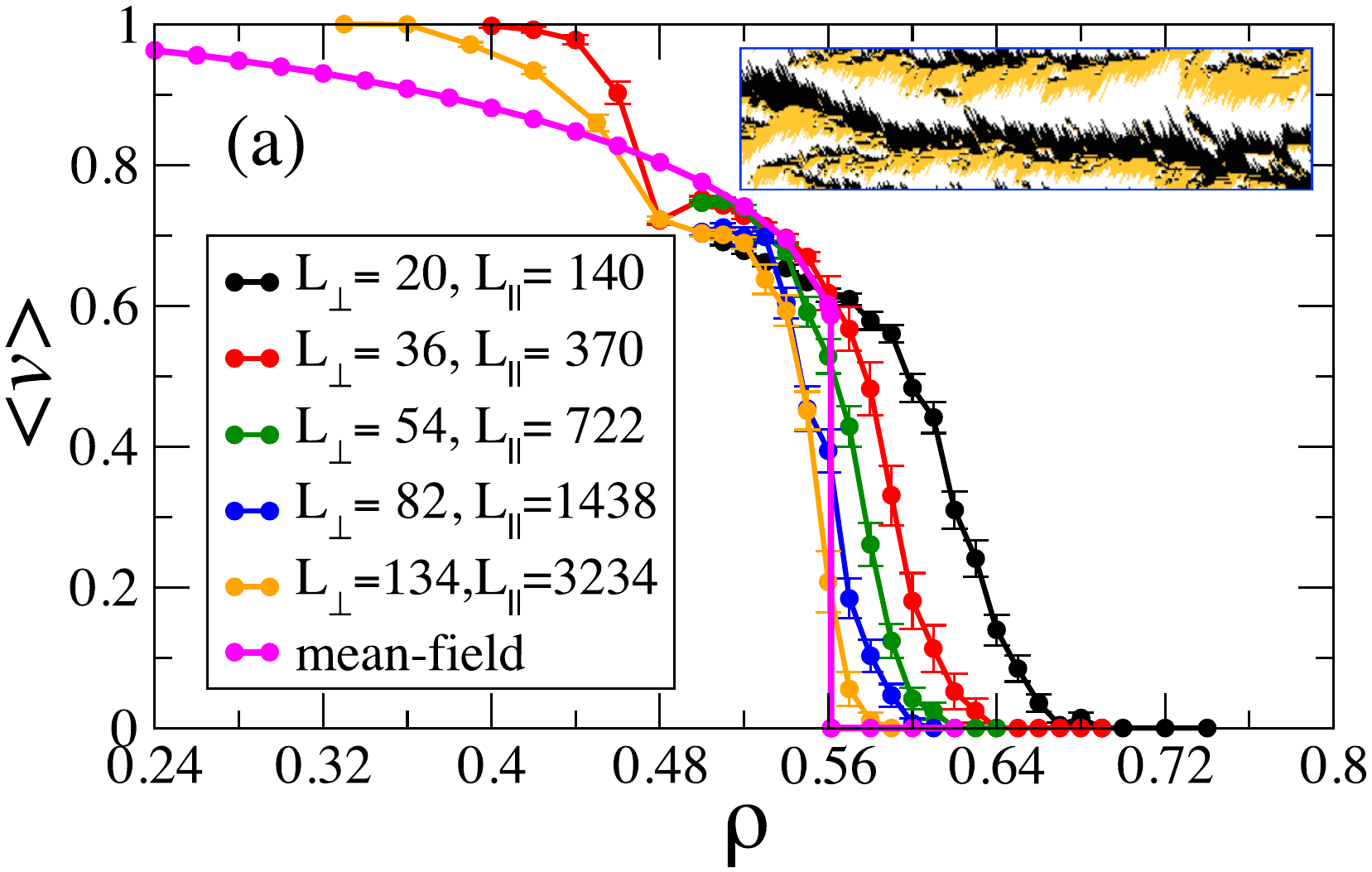}\quad
   \includegraphics[width=0.35\textwidth]{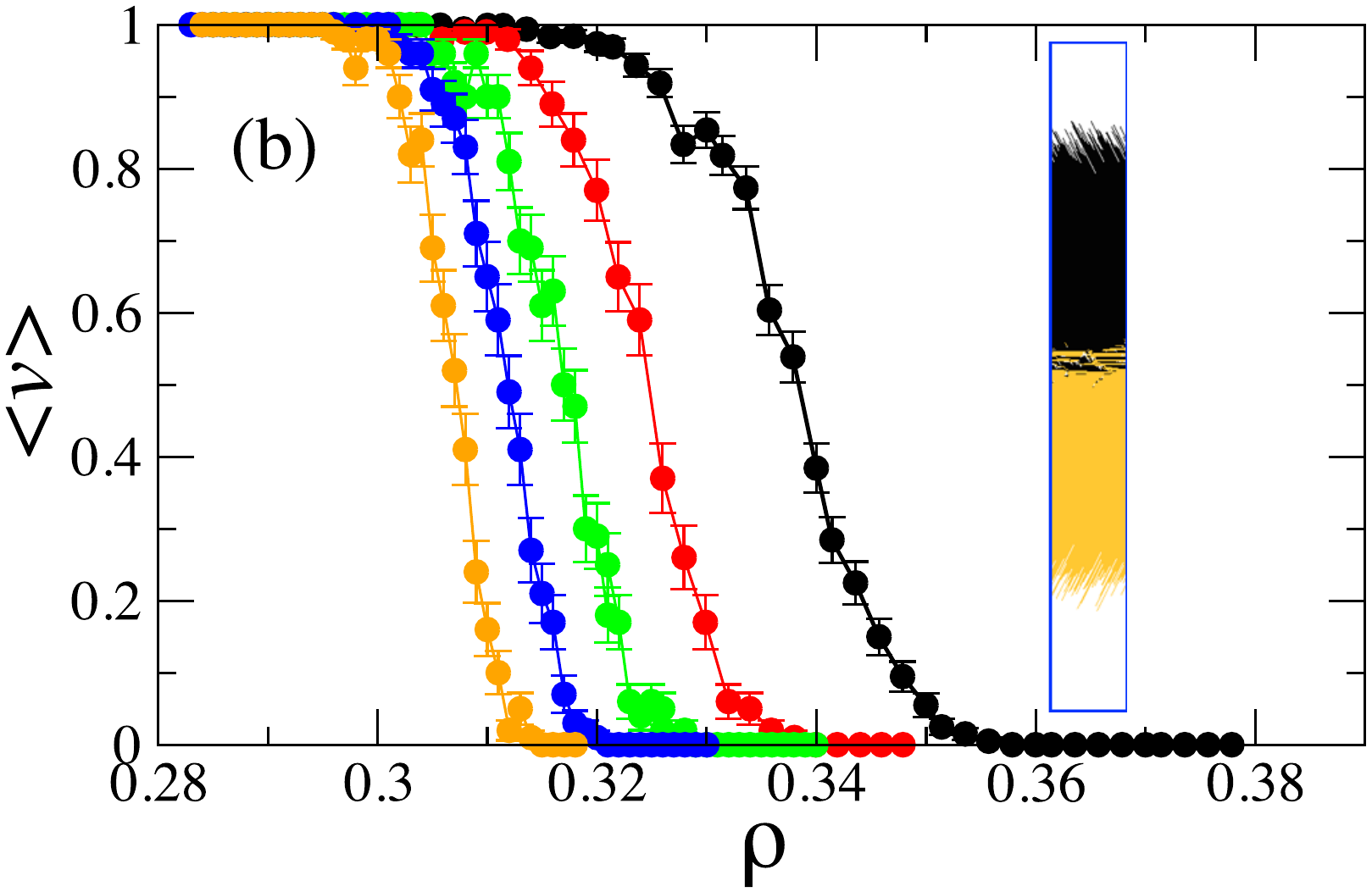}
   \includegraphics[width=0.36\textwidth]{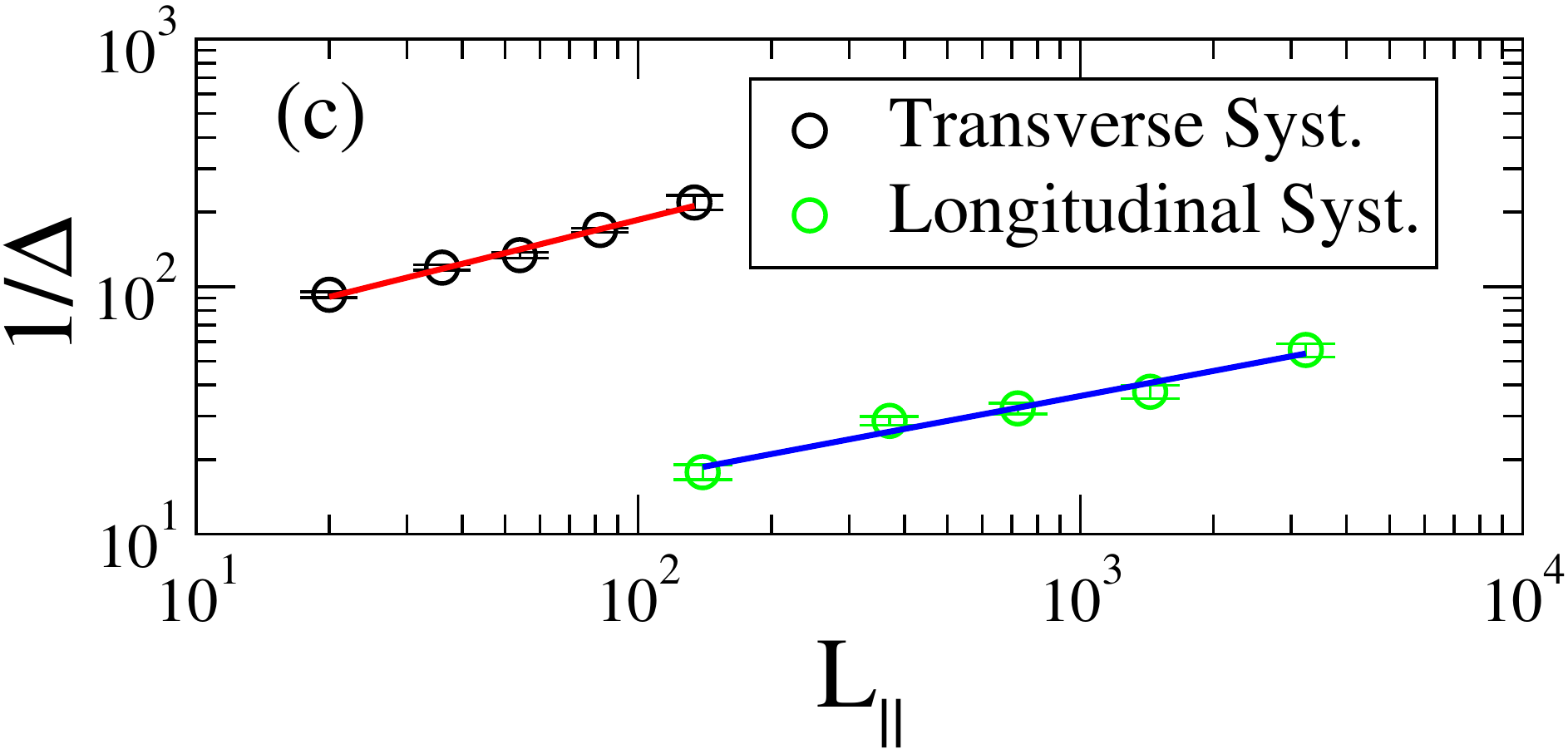}\quad
  \includegraphics[width=0.36\textwidth]{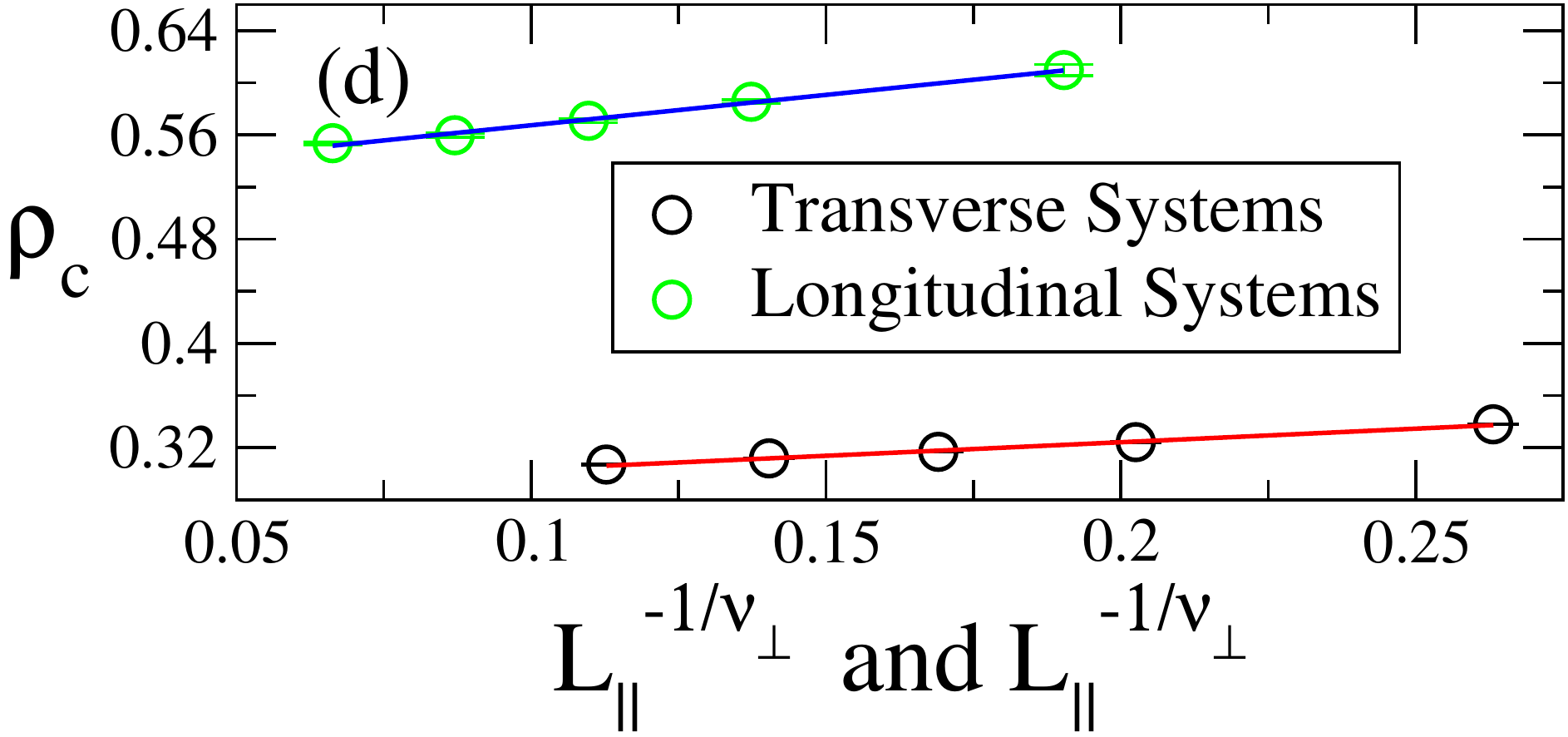}
\caption{(color online). Finite-size scaling analysis for the phase transition of both longitudinal and transversal systems. (a) Transition curves for several longitudinal systems, including the mean-field prediction (magenta) and a typical configuration for the jammed phase (inset).  (b) Transition curves for several transversal systems (sizes as in (a), but with $L_\parallel$ and  $L_\perp$ interchanged).  (c) Scaling of $\Delta^{-1}(L_\parallel)$ for both cases, giving $\nu_{\parallel}$$=$$3.0(3)$ and  $\nu_{\perp}$$=$$1.8(2)$. (d) Scaling of the finite critical densities give $\rho_{c_\parallel}$$=$$0.521(2)$ and  $\rho_{c_\perp}$$=$$0.283(2)$.
All systems were simulated for at least $1.5 \times10^6$ time steps or until convergence ($v$$=$$0$ or $v$$=$$1$).}
\label{allfss}
\end{figure*}
Figure \ref{allfss}(a) shows that velocity begins to decrease smoothly with increasing $\rho$ until the abrupt onset of full jamming ($v$$=$$0$) at a certain $\rho_{c_\parallel}$. Right there, an approximately uniform distribution of jams spans through the whole system (Fig.\ref{allfss}(a), inset).  With Eq.(\ref{scale}) in mind, the FSS analysis (Fig.\ref{allfss}(c) and (d)) gives  $\rho_{c_\parallel}$$=$$0.521(2)$ and $\nu_{\parallel}$$=$$3.0(3)$, which would imply  $\nu_{\perp}$$=$$1.8(2)$. 

Quite differently, transversal systems (i.e., lattices with  $L_{\parallel}\times L_{\parallel}^{\frac{1}{\theta}}$) in the gridlock phase show a single and well localized jam on an empty background (Fig.\ref{allfss}(b)). These systems exhibit a sharp phase transition between free-flow and a completely gridlock phases, but at a lower density $\rho_{c_\perp}$. The FSS analysis (also  in Fig.\ref{allfss}(c) and (d)) gives  $\rho_{c_\perp}$$=$$0.283(1)$ and $\nu_{\perp}$$=$$2.2(1)$, in agreement with the previous result. Hence, the intermediate states in square lattices \cite{raissa1} emerge just as a consequence of the combination of these two phase transitions (Fig.\ref{allfss}(a) and (b)). 

The critical densities for both longitudinal and traverse transitions can be approximated by using a mean-field analysis, inspired by \cite{wangimproved}. Consider the mean velocity of brown cars (by symmetry, the reasoning is also valid for black cars). A brown car will stop either because it is crossed by a black car or because it queues behind another brown car. At a random initial configuration, the probability that a car is crossed or queued  is $\rho^2$, that is, at the beginning of the simulation the proportion of stopped cars $p_{\rm stop}$ must be equal to $\rho$. 
Let us define $c_{\to\uparrow}$ ($c_{\to\to}$) as the proportion of stopped cars that are crossed (queued). In previous works \cite{wangimproved,nagatanimean}, it has been assume that  $c_{\to\uparrow}$$=$$c_{\to\to}$$=$$0.5$, but we will see that this is not the case. If $p_{\rm stop}\sim\rho$ for some time steps, the probability a cell to be occupied by a stopped crossed (queued) car will be $c_{\to\uparrow}\rho^2$($c_{\to\to}\rho^2$).
Since black cars spend on average a time $1/v$ on a site, they will reduce the speed of brown cars from unity by $c_{\to\uparrow}\rho^2/v$. Similarly, the extra amount of time that a brown car stays on a site will be given by $\frac{1}{v}$$-$$1$, reducing the average speed by  $c_{\to\to}\rho^2(\frac{1}{v}-1)$. Hence, a self-consistency equation for the average speed $v$ will be
\begin{equation}\label{meanfield}
v=1-\frac{c_{\to\uparrow}\rho^2}{v}-c_{\to\to}\rho^2\left(\frac{1}{v}-1\right),
\end{equation}
which gives $\rho_c$ as the critical density at which the equation ceases to give a real solution. 
\begin{figure}[t]
\centering
\includegraphics[width=0.27\textwidth]{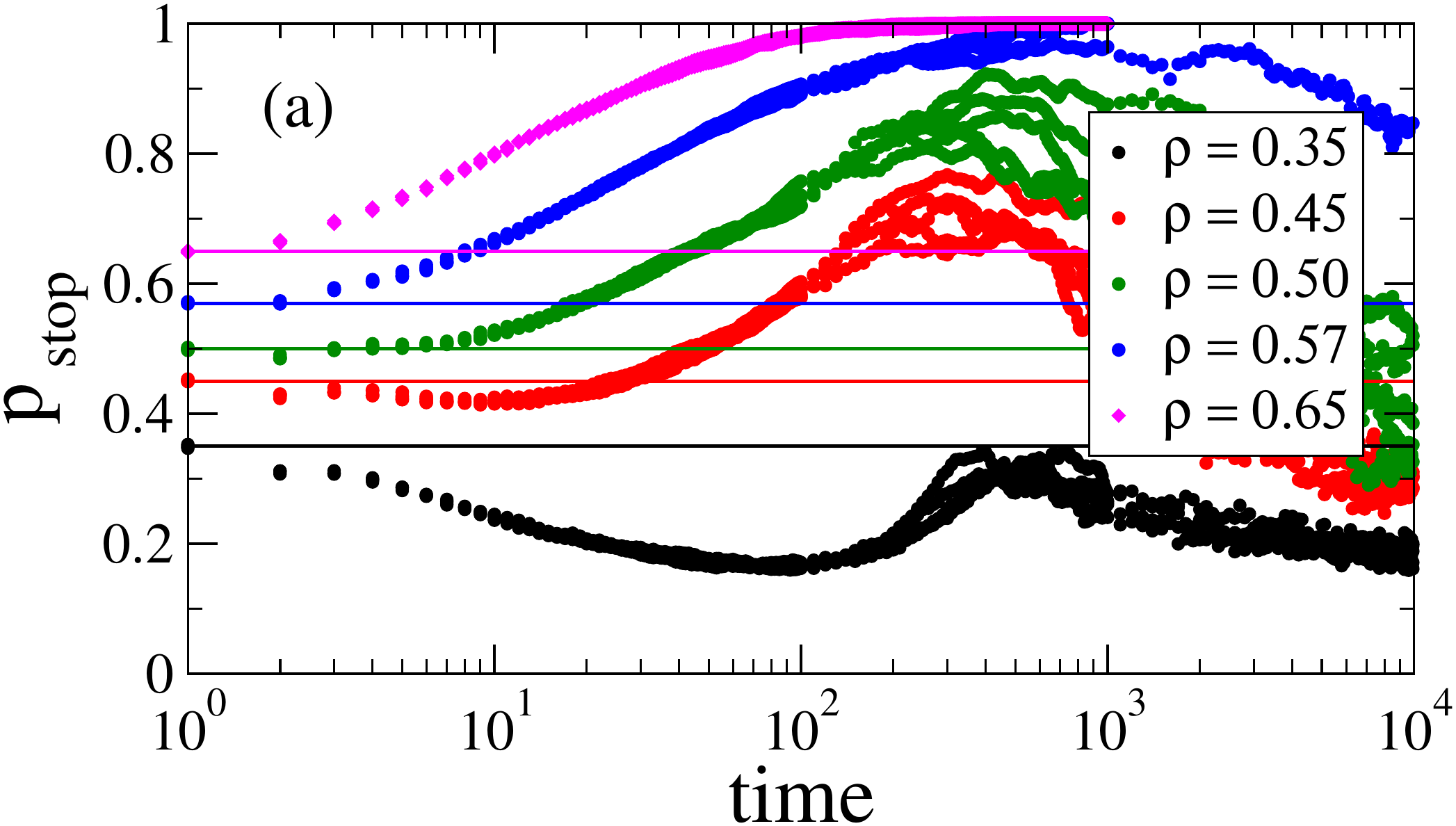}\quad
\includegraphics[width=0.17\textwidth]{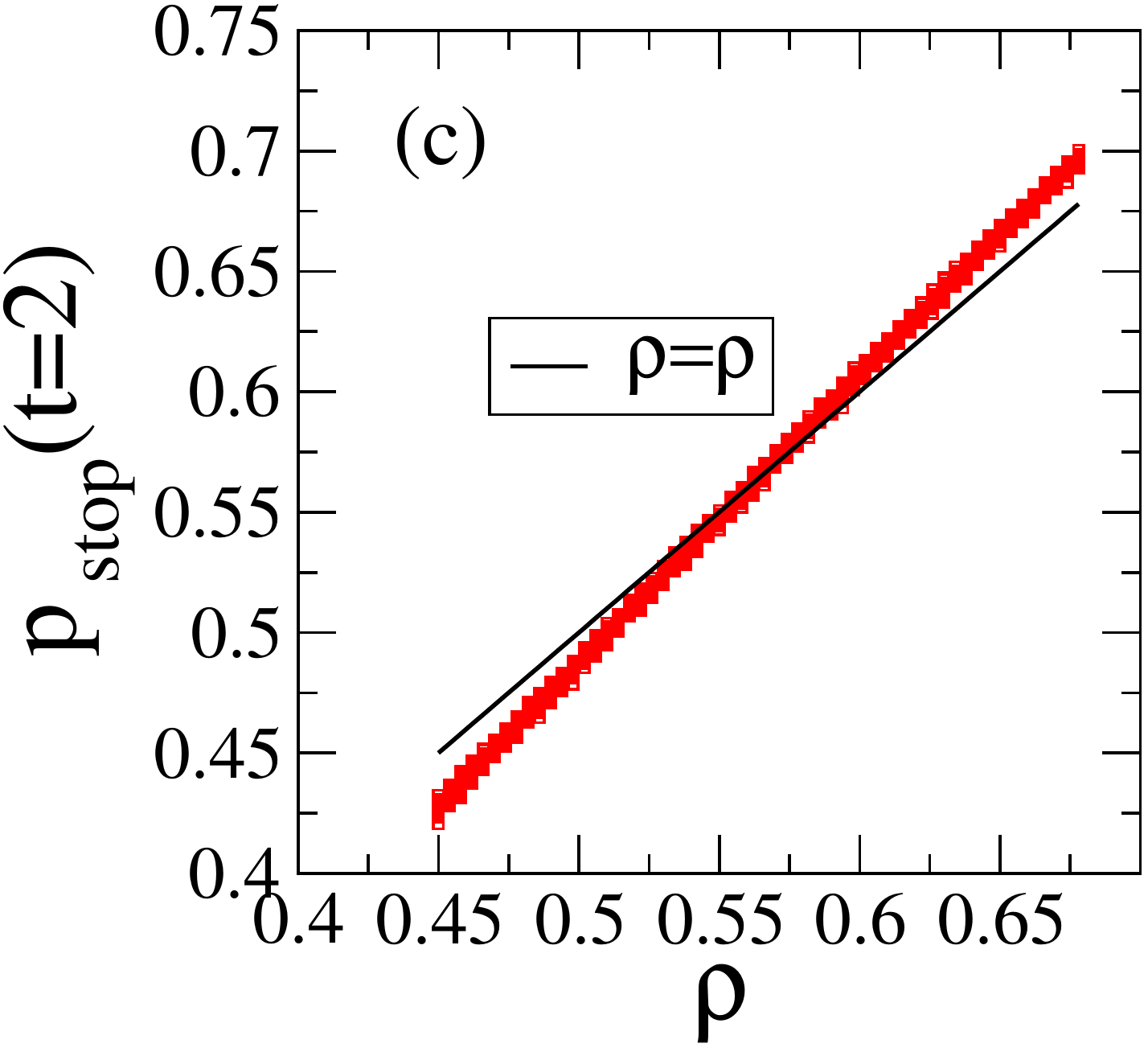}\\
\includegraphics[width=0.27\textwidth]{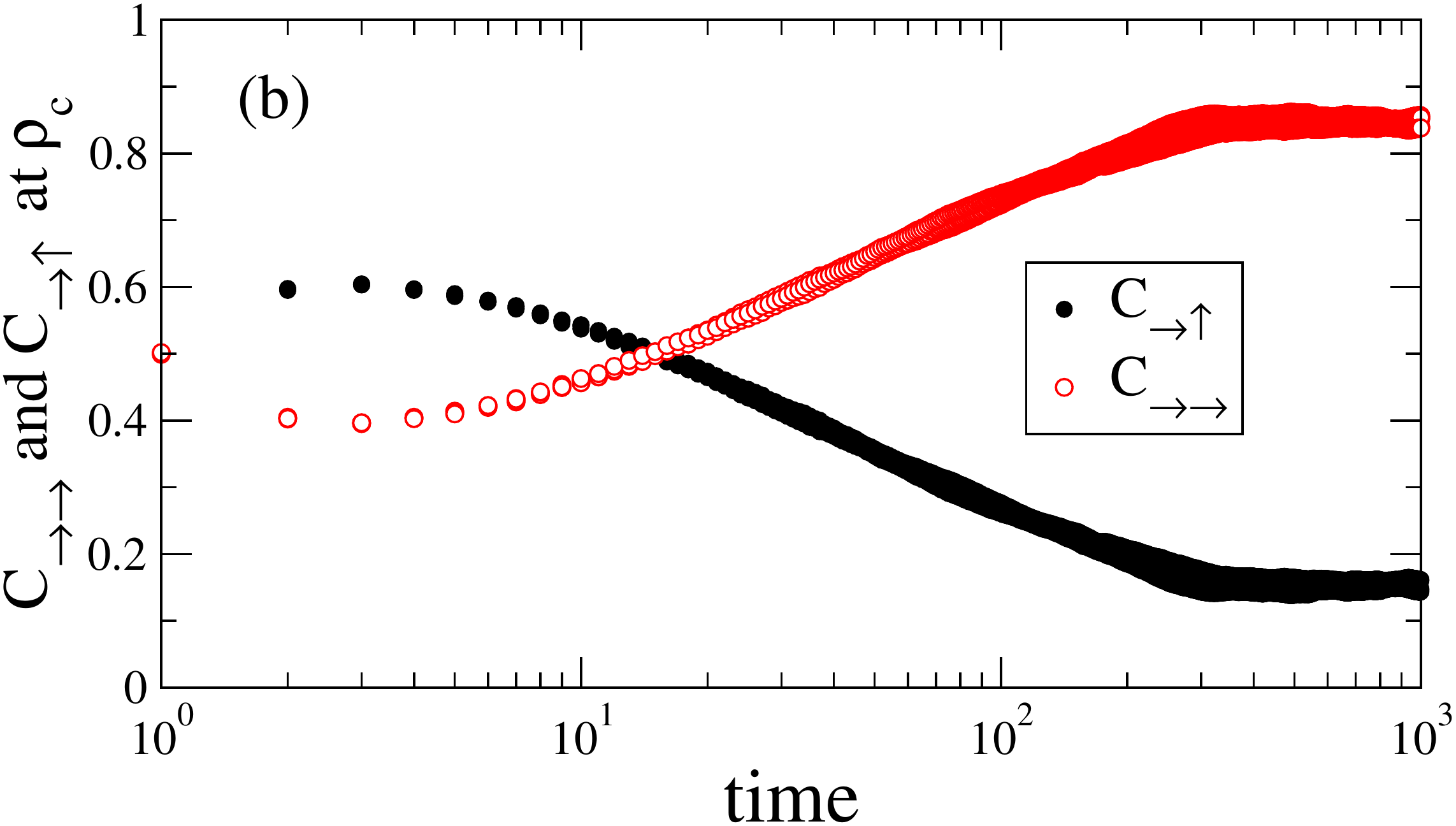}\quad
\includegraphics[width=0.17\textwidth]{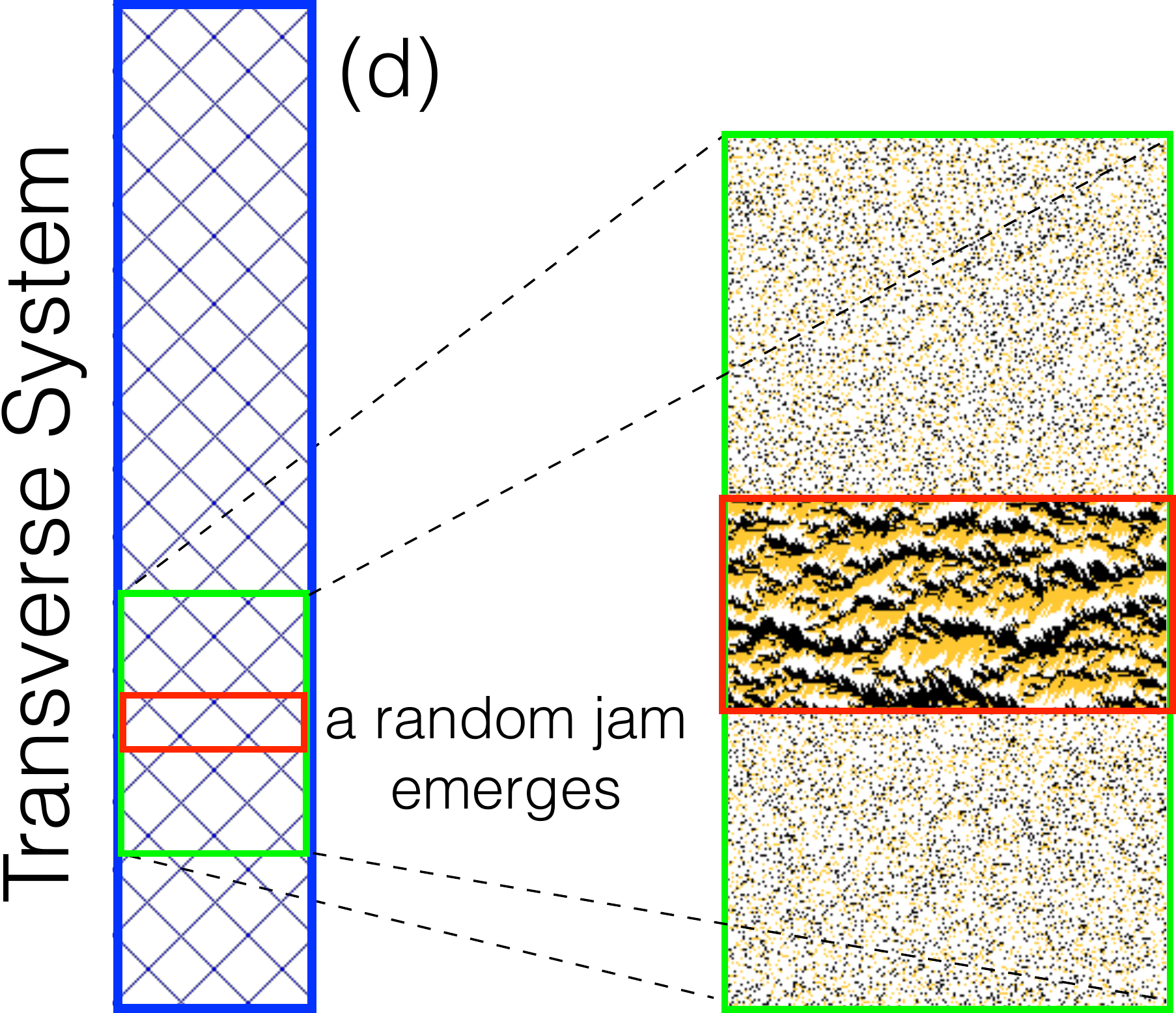}
\caption{(color online). Time evolution of the fraction of stopped cars $p_{\rm stop}$ (a) and of the fraction of crossed ($c_{\to\uparrow}$) and queued ($c_{\to\to}$) stopped cars (b) in a $1438\times 82$ longitudinal system. The value of $p_{\rm stop}$ after one time step for several densities is also shown (c). (d) Schematic representation of the jamming process for transversal systems}\label{cross}
\end{figure}

Consider  longitudinal systems first, where a uniform distribution of longitudinal jams arise. Before solving Eq.(\ref{meanfield}), let us study the evolution on time of the quantities involved. On one hand, Fig.\ref{cross}(a) shows the time evolution of $p_{\rm stop}$ for different values of the car density for a longitudinal system. At start, $p_{\rm stop}$$=$$\rho$ for every density, as expected. Later on, $p_{\rm stop}$  declines (grows) for low (high) densities, and the crossover occurs near the critical density $\rho_c$. Actually, $p_{\rm stop}$ after one time step equals $\rho$ for $\rho\simeq 0.57$ in a large system (Fig.\ref{cross}(b)). On the other hand, $c_{\to\uparrow}$$=$$c_{\to\to}$$=$$0.5$ only at start, but after just one time step they change to $c_{\to\uparrow}$$\sim$$0.600(5)$  and $c_{\to\to}$$\sim$$0.400(5)$, remaining there for some time steps (Fig.\ref{cross}(c)).  By replacing these two values into Eq.(\ref{meanfield}), the mean-field approach predicts a critical density $\rho_\parallel$$=$$0.563(5)$, in agreement with the value  obtained from both finite-size scaling and the crossover described above.

Transversal systems, in contrast, collapse into a single big jam. Thus, the previous reasoning is impractical here, because the local density becomes inhomogeneous during the collapse: increasing in the vicinity of the growing jam and dropping to zero elsewhere. Let us consider that the density at start is low enough to assume that all cars are moving. At a certain evolution time the cars have condensed into a thinner region (red area in Fig.\ref{cross}(d)) of area $A_{\rm red}$, where the density has reached the critical value $\rho_{c_\parallel}$ for longitudinal systems. All cars in this region come from a larger region of area $A_{\rm green}$ in the initial configuration (green area in Fig.\ref{cross}(d)). Let us assume that the red area is a fraction of the green one equals to the fraction of cars that have stopped, $A_{\rm red}/A_{\rm green}$$=$$p_{\rm stop}$. Because the number of cars does not change, at the critical density
\begin{equation}
\rho_{green}=\rho_{red} \frac{A_{red}}{A_{green}}=\rho_{c_\parallel}^2\quad,
\end{equation}
where we have used the result that $p_{\rm stop}=\rho$ at the critical density for longitudinal systems. Then, the critical density for transversal systems will be $\rho_{c_\perp}$$=$$\rho_{c_\parallel}^2$$=$$0.271(3)$, which approximates quite well the value $\rho_{c_\perp}$$=$$0.283(2)$ from the simulations (Fig.\ref{allfss}(d)). 

The values for $c_{\to\to}$ and $c_{\to\uparrow}$ can also be  estimated from the scaling analysis. Consider a typical jam (as the one in Fig.\ref{allfss}(a)). It consists of two sets of queues, east-bound (brown) and north-bound (black), intersecting along a longitudinal curve. Crossed cars localize on this intersection, and its number is proportional to $L_\parallel$$\propto$$L_\perp^\theta$. Queued cars fill the whole jam, and its number is proportional to the jam's area (i.e. to $L_\parallel \times L_\perp$). Thus, $c_{\to\uparrow}$$\propto$$L_\perp^\theta$ and $c_{\to\to}$$\propto$$L_\perp^{\theta+1}$$\propto$$c_{\to\uparrow}^\frac{\theta+1}{\theta}$. Replacing into $c_{\to\to}+c_{\to\uparrow}$$=$$1$, they give $c_{\to\uparrow}$$=$$0.582(1)$ and $c_{\to\to}$$=$$0.418(1)$, very close to the simulation results. Using these values in the mean-field equation gives $\rho_\parallel$$=$$0.5672(2)$, also close to the simulations.

Taken together, our results suggest that the puzzling intermediate states in the BML model on square lattices are, actually, a consequence of the system's anisotropy, which produces two different phase transitions: one for transverse systems, with $\rho_{c_\perp}$$=$$0.283(1)$, and another for longitudinal ones, with  $\rho_{c_\perp}$$=$$0.521(1)$. Indeed, the first critical density corresponds with the lower bifurcation point on square lattices, reported at $\rho$$=$$0.315$ \cite{raissa1}, contradicting the general believe that this would go to zero for infinite systems; similarly, the second critical density perfectly matches with the value of  $\rho$$=$$0.52$ reported as the transition point between self-organized jams and random jams \cite{tadaki,gupta} within the conventional understanding of the  original BML model. The asymptotic limit of the curves from simulations on transverse systems (blue line in Fig.\ref{FigIntermediateStates}) and the mean-field curve for longitudinal systems (Eq. (\ref{meanfield}), red line in Fig.\ref{FigIntermediateStates}) perfectly enclose the zone of intermediate states.

Despite having studied the critical features of the BML phase transitions with the DP formalism, the obtained critical exponents are not compatible with the universality class of the directed percolation. Instead, the critical behavior here coincides with the reported for the parity-conserving universality class (PC)\cite{hinrich,nonequi}. The origin of this relationship remains as an open question for future research.  

By finding the origin of the intermediate states of the BML model, we have built a very complete description of such a fundamental model for traffic flow, illustrating at the same time the power of the finite-size scaling analysis for anisotropic systems, where the BML model seems to be a paradigmatic example. We look forward to see many more results of this enlightening analysis technique in the future.
\begin{acknowledgments}
We thank the Centro de Estudios Interdisciplinarios B\'asicos y Aplicados en Complejidad, CeiBA-Complejidad, and the Universidad Nacional de Colombia for financial support.
\end{acknowledgments}

\bibliography{apssamp}

\end{document}